\documentclass[12pt, 1p, review, number]{elsarticle}
\usepackage[utf8]{inputenc}
\usepackage[bitstream-charter]{mathdesign}
\usepackage[T1]{fontenc}
\usepackage{psfrag}
\usepackage{subfigure}
\usepackage{array}
\usepackage{multirow}
\usepackage{hhline}
\usepackage{color}
\usepackage{setspace}
\usepackage{amsmath}
\usepackage[pagewise,modulo]{lineno}
\linenumbers
\usepackage{pifont}
\usepackage[pagebackref,colorlinks=True]{hyperref}
\usepackage{wasysym}
\usepackage{color}
\DeclareTextSymbol{\degr}{T1}{6}
\newcommand{\degre}[0]{\degr C}
\makeatletter
 \renewcommand\@journal{Acta Materialia}
\makeatother
\begin{document}
%------------------------------
% Presentation
%------------------------------
\begin{frontmatter}
\title{A method for measuring the contact area in instrumented indentation testing by tip scanning probe microscopy imaging}
\author[Annecy,Rennes]{L. Charleux\corref{cor1}} 
\author[Lorient,Rennes]{V. Keryvin} 
\author[Rennes]{M. Nivard}
\author[Rennes]{J.-P. Guin}
\author[Rennes]{J.-C. Sangleb\oe uf}
\author[Sendai]{Y. Yokoyama}
\address[Annecy]{Univ. Savoie, EA 4144, SYMME, F-74940 Annecy-le-Vieux, France}
\address[Lorient]{Univ. Bretagne-Sud, EA 4250, LIMATB, F-56100 Lorient, France}
\address[Rennes]{Univ. Rennes 1, ERL CNRS 6274, LARMAUR , F-35000 Rennes, France}
\address[Sendai]{Institute for Materials Research, Tohoku University, Sendai 980-8577, Japan}

\cortext[cor1]{Corresponding author: ludovic.charleux@univ-savoie.fr}

%------------------------------
% Abstract
%------------------------------
\begin{abstract}
The determination of the contact area is a key step to derive mechanical properties such as hardness or an elastic modulus by instrumented indentation testing. Two families of procedures are dedicated to extracting this area: on the one hand, \textit{post mortem} measurements that require residual imprint imaging, and on the other hand, direct methods that only rely on the load \textit{vs.} the penetration depth curve. With the development of built-in scanning probe microscopy imaging capabilities such as atomic force microscopy and indentation tip scanning probe microscopy, last generation indentation devices have made systematic residual imprint imaging much faster and more reliable. In this paper, a new \textit{post mortem} method is introduced and further compared to three existing classical direct methods by means of a numerical and experimental benchmark covering a large range of materials. It is shown that the new method systematically leads to lower error levels regardless of the type of material. Pros and cons of the new method \textit{vs.} direct methods are also discussed, demonstrating its efficiency in easily extracting mechanical properties with an enhanced confidence. 

%\noindent[PLEASE DISPLAY FIGURE 4 WITH THE ONLINE ABSTRACT]

\end{abstract}

%------------------------------
% Chosen figure for the online abstract
%------------------------------

%------------------------------
% Keywords
%------------------------------
\end{frontmatter}

\section{Introduction}

Over the last two decades, Instrumented Indentation Technique (IIT) has become a widespread procedure that is used to probe mechanical properties for samples of nearly any size or nature. However, the intrinsic heterogeneity of the mechanical fields underneath the indenter prevents from establishing straightforward relationships between the measured load \textit{vs.} displacement curve and any expected mechanical properties as it would be the case for a tensile test. Many models have been published in the literature in order to enable the measurement of properties such as an elastic modulus, hardness or various plastic properties. Despite their diversity, most of these models deeply rely on the accurate measurement of the projected contact area between the indenter and the sample's surface. The existing methods that are dedicated to estimating the true contact area can be classified into two subcategories: the direct methods which rely on the sole load \textit{vs.} displacement curve \cite{JMR:7931088, ISI:000222316100002, ISI:A1984SQ59800006} and the \textit{post mortem} methods that use additional data extracted from the residual imprint left on the sample's surface. For example, Vickers, Brinell and Knoop hardness scales rely on \textit{post mortem} measurements of the geometric size of the residual imprint. However, in the case of Vickers hardness, the contact area is only estimated through the diagonals of the imprint, the possible effect of piling-up or sinking-in is then neglected. Other \textit{post mortem} methods use indent cross sections to estimate the projected contact area \cite{Beegan2004,Zhou2008}. In the 1990s, the development  of nanoindentation led to a growing interest in direct methods because they do not require time consuming \textit{post mortem} measurement of micrometer or even nanometer scale imprints, typically using Atomic Force Microscopy (AFM) or Scanning Electron Microscopy (SEM). Uncertainty level on  direct measurements remains high, mainly because of the difficulty to predict the occurrence of piling-up and sinking-in. Oliver and Pharr have eventually considered this issue as one of the \textit{"holy grails"} in IIT \cite{ISI:000222316100002}.  Recent development in Scanning Probe Microscopy (SPM) using the Indentation Tip (ITSPM) brought new interest in \textit{post mortem} measurements. Indeed, ITSPM allows systematic imprint imaging without manipulating the sample or facing repositioning issues to find back the imprint to be imaged. Yet ITSPM imaging technique suffers from drawbacks when compared to AFM: it is slower, it uses a blunter tip associated with a much wider pyramidal geometry and a higher force applied to the surface while scanning. While the later may damage delicate material surfaces, the formers will introduce artifacts. Nonetheless, these artifacts will not affect the present method. In addition, ITSPM only allows for contact mode imaging, non contact or intermittent contact modes are not possible. As a consequence, only the techniques based on altitude images can be used with ITSPM and there is a need for new methods as very recently reviewed by Marteau \textit{et al.} \cite{Marteau2013}. This article introduces a new \textit{post mortem} procedure that relies only on the altitude image and that is therefore valid for most types of SPM images, including ITSPM. In this paper, a benchmark based on both numerical indentation tests as well as experimental indentation tests on properly chosen materials to span all possible behaviors is first introduced. Then, the existing direct methods are reviewed and a complete description of the proposed method is given. These methods are then confronted using the above mentioned benchmark and the results are finally discussed.
 
%The method is applied to both numerical tests using the Finite Element Method (FEM) and experimental tests on 3 different glasses exhibiting a wide range of contact behaviors, from sinking-in to piling-up. The model is also confronted the 3 main direct methods available in the literature. 

\section{Numerical and experimental benchmark}

\begin{figure}[p]
\begin{center}
\includegraphics[width = .8\textwidth]{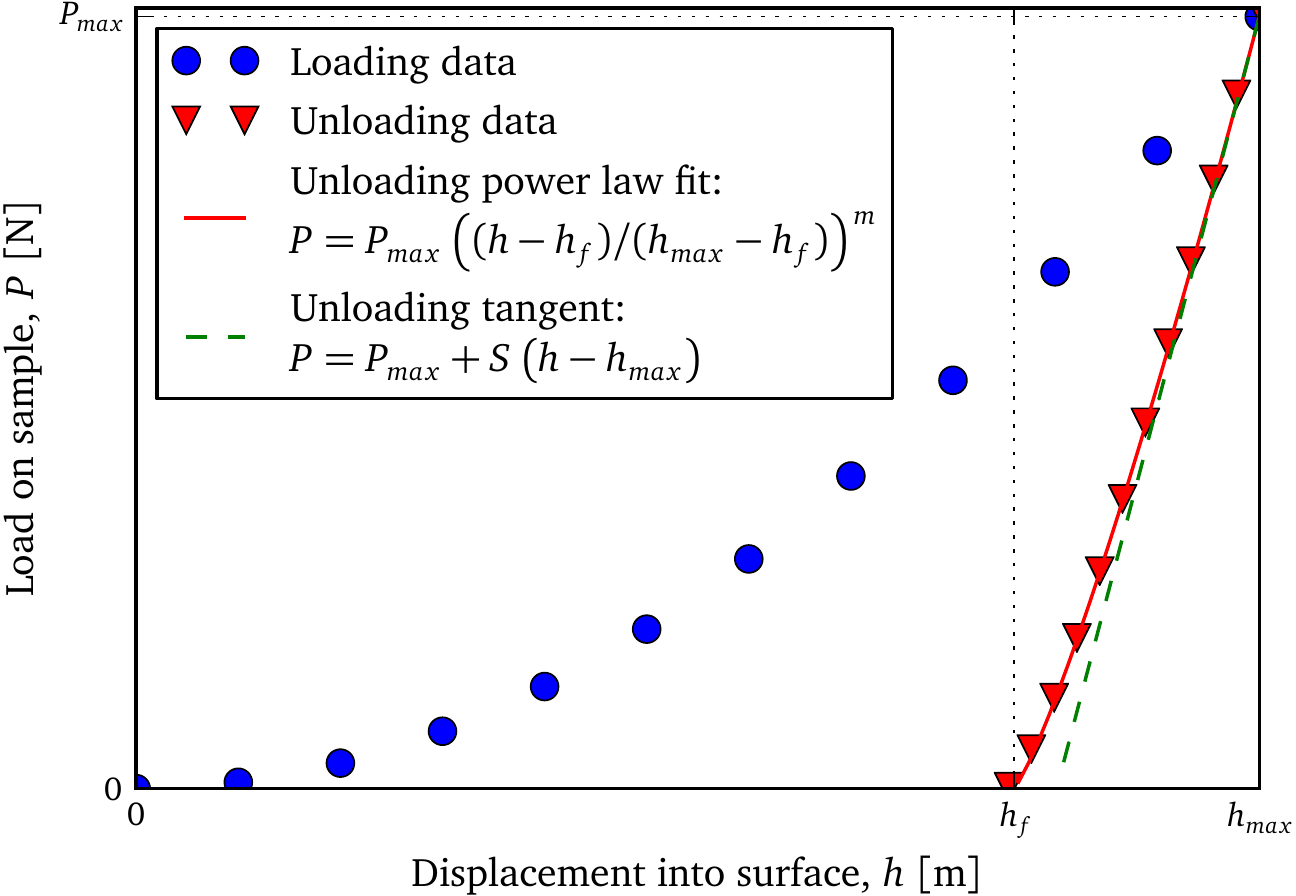}
\caption{Typical sharp indentation load on sample \textit{vs.} displacement into surface curve. The test is split into a loading step and an unloading step. The experimental curve generally also includes an holding step which is not represented in this case. The contact stiffness $S$ is the unloading step's initial slope. However, the direct determination of $S$ \textit{via} the upper part of the step is unreliable as it uses only a small part of the curve. For increased accuracy, the whole step is systematically fitted by a power law function which is used to compute back the contact stiffness $S$ as initially recommended in \cite{ISI:000222316100002}.}
\label{fig:figure_1}
\end{center}
\end{figure}

A typical instrumented indentation test features a loading step where the load $P$ is increased up to a maximum value $P_{max}$, then held constant in order to detect creep and finally decreased during the unloading step until contact is lost between the indenter and the sample. A residual imprint is left on the initially flat surface of the sample. During the test, the load $P$ as well as the penetration of the indenter into the surface of the sample $h$ is continuously recorded and can be plotted as shown in Figure \ref{fig:figure_1}. For most materials, the unloading step can be cycled with only minor hysteresis, it is then assumed that only elastic strains develop in the sample. As a consequence, the initial slope $S$ of the  unloading step is called the elastic contact stiffness. Useful data can potentially be extracted from both the load \textit{vs.} displacement curve and the residual imprint. The contact area $A_c$ is defined as the projection of the contact zone between the indenter and the sample at maximum load on the plane of the initially flat surface of the sample.

\subsection{Numerical approach}

\begin{figure}[p]
\begin{center}
\includegraphics[width = 1.\textwidth]{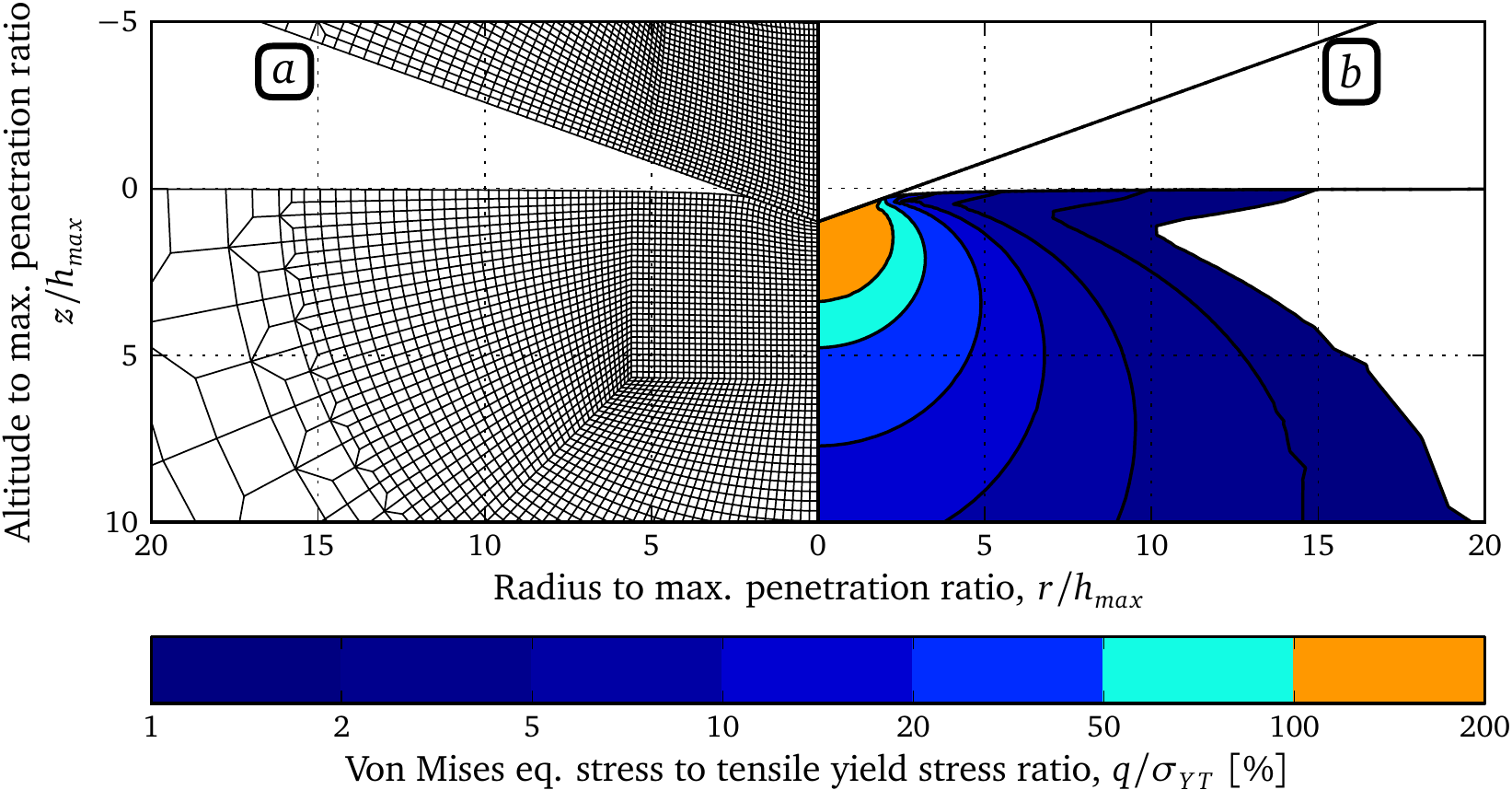}
\caption{Representation of the 2D axisymmetric FEM model including a rigid conical indenter and a deformable elastic-plastic sample. In this particular case, the sample's material is based on CE1 ($\sigma_{YT}/E = 0.07$, $n = 0.3$). The model is represented at maximum penetration into surface $h_{max}$.  (\textbf{a}) The deformed mesh is plotted showing the finely meshed zone containing the contact zone and most of the plastic zone. This zone is initially filled with square shaped elements. The mesh is gradually coarsened away from the contact zone. The whole mesh is not represented since its total size is about $10^3$ times the size of the finely mesh zone. (\textbf{b}) The gradient represents the ratio of the von Mises equivalent stress in tension $q$ to the tensile yield stress $\sigma_{YT}$ with the corresponding scale given by the bottom color bar.}
\label{fig:figure_2}
\end{center}
\end{figure}

Finite element modeling (FEM) simulations are performed using a two-dimensional axisymmetrical model represented in Figure \ref{fig:figure_2}. The sample is meshed with 3316 four-noded quadrilateral elements. The indenter is considered as a rigid cone exhibiting an half-angle $\Psi = 70.29$\degr~to match the theoretical area function of the Vickers and modified Berkovich indenters \cite{Fischer-Cripps2011}. The displacement of the indenter $h$ is controlled and the force $P$ is recorded. The dimensions of the mesh are chosen to minimize the effect of the far-field boundary conditions. The typical ratio of the maximum contact radius and the sample size is about $2\times10^3$. The problem is solved using the commercial software ABAQUS (version 6.11,~\href{3ds.com}{3ds.com}). The numerical model is compared to the elastic solution from \cite{love} (see \cite{Hanson1992, Sneddon1965}) using a blunt conical indenter ($\Psi = 89.5$\degr) to respect the purely axial contact pressure hypothesis used in the elastic solution. The relative error is computed from the load \textit{vs.} penetration curve and is below $0.1\%$. Pre-processing, post-processing and data storage tasks are performed using a dedicated framework based on the open source programming language Python 2.7 \cite{VanRossum1991, Hunter2007, Oliphant2007} and the database engine SQLite 3.7 \cite{Owens2006}. The indented material is assumed to be isotropic, linearly elastic. The Poisson's ratio $\nu$ has a fixed value of $0.3$ and the Young's modulus is referred to as $E$. The contact between the indenter and the sample's surface is taken as frictionless. Two sets of constitutive equations (CE1 and CE2) are investigated in order to cover a very wide range of contact geometries and materials: 

\begin{description}
\item[CE1] This first constitutive equation used in this benchmark is commonly used in industrial studies and in research papers on metallic alloys \cite{Bucaille2003, Dao2001, Ma2011, ISI:000235223500006}. It uses $J_2$-type associated plasticity and an isotropic Hollomon power law strain hardening driven by the tensile behavior (stress $\sigma_T$, strain $\epsilon_T$) given by Eq.  \ref{eq:hollomon}:
\begin{equation}
\sigma_T = \left\lbrace
\begin{split}
E \epsilon_T, \qquad \mbox{for: } \sigma_T \leq \sigma_{YT}\\
\sigma_{YT}\left( \dfrac{E \epsilon_T}{\sigma_{YT}}\right)^n,\qquad \mbox{for: } \sigma_T > \sigma_{YT}
\end{split}\right.
\label{eq:hollomon}
\end{equation}
Plastic parameters are the tensile yield stress $\sigma_{YT}$ and the strain hardening exponent $n$.

\item[CE2] The second constitutive equation is the Drucker-Prager law \cite{Drucker1952} which was originally dedicated to soil mechanics but was also found to be relevant on Bulk Metallic Glasses (BMGs)\cite{ISI:000254173700030,Patnaik2004,ISI:000247980200005,Keryvin2007} and some polymers \cite{Prasad2009}. The yield surface is given by Eq. \ref{eq:drucker-prager} where $q$ is the von Mises equivalent stress in tension and $p$ the hydrostatic pressure. Perfect plasticity is used in conjunction with an associated plastic flow. The plastic behavior is controlled by the compressive yield stress $\sigma_{YC}$ and the friction angle $\beta$ that tunes the pressure sensitivity. 

\begin{equation}
q - p \tan \beta - \left(1 - 1/3 \tan \beta \right) \sigma_{YC} = 0
\label{eq:drucker-prager}
\end{equation}

\end{description}

Dimensional analysis \cite{buckingham,Buckingham1915, ISI:000075294600019} is used to determine the influence of elastic and plastic parameters on the contact area $A_c$:

\begin{equation}
A_c = \left\lbrace\begin{split}
h_{max}^2 \Pi_{CE1}(\nu,\sigma_{YT}/E, n) \\
h_{max}^2 \Pi_{CE2}(\nu,\sigma_{YC}/E, \beta) 
\end{split}\right.
\end{equation} 

In this equation, $h_{max}$ is the maximum value of penetration of the indenter into the sample's surface. In both cases, the dimensionless functions show that, since the Poisson's ratio has a fixed value ($\nu = 0.3$), only the yield strains ($\epsilon_{YT} = \sigma_{YT}/E$, in the case of CE1 and $\epsilon_{YC} = \sigma_{YC}/E$, in the case of CE2) and the dimensionless plastic parameters ($n$ in the case of CE1 and $\beta$ in the case of CE2) have an influence on the contact area $A_c$. As a consequence, the value of the Young's modulus $E$ has a fixed arbitrary value $E = 1$ Pa and only the values of the yield stresses $\sigma_{YT}$ and $\sigma_{YC}$, the hardening exponent $n$ and the friction angle $\beta$ are modified. The simulated range of these parameters are given in Tables \ref{tab:hollomon_params} and \ref{tab:dp_params}. After each simulation, a load \textit{vs.} displacement into surface curve and an altitude SPM like image using the Gwyddion (\href{http://gwyddion.net/}{http://gwyddion.net/}) GSF format are extracted. The use of such a procedure allows one to consider both numerical and experimental tests in the benchmark and to derive mechanical properties in the same way. Since the simulations are two-dimensional axisymmetric, the contact area is computed as $A_c = \pi r_c^2$ where $r_c$ stands for the contact radius of the contact zone (see Fig. \ref{fig:figure_2}).

\subsection{Experimental testing}
\label{subsec:expe_testing}

\begin{figure}[p]
\begin{center}
\includegraphics[width = 1.\textwidth]{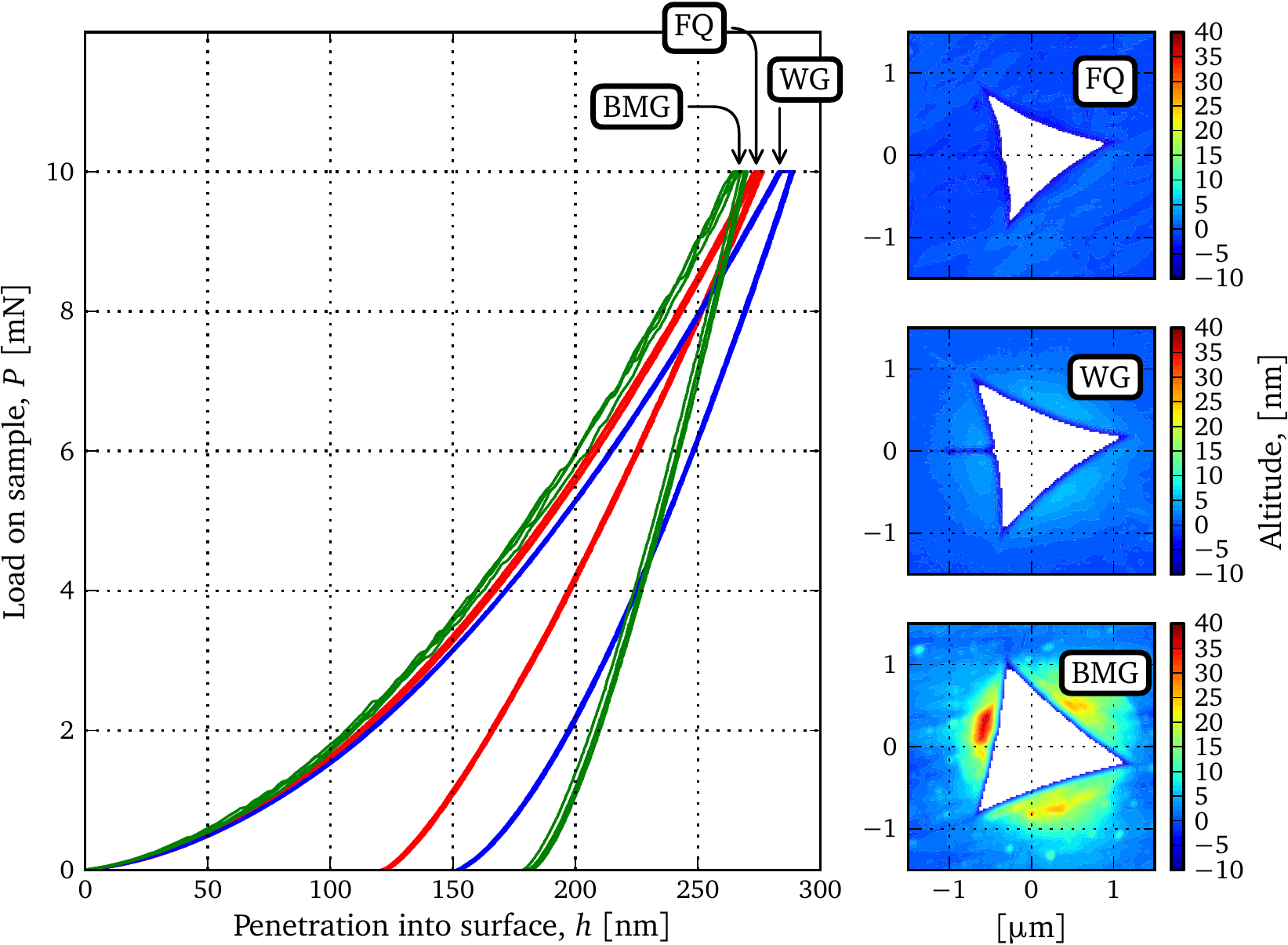}
\caption{Representation of the experimental tests carried out on the three selected materials FQ, WG and BMG (Four tests per sample). (\textbf{Left}) The load \textit{vs.} displacement into surface $(P,h)$ curves are plotted. The three loading steps are rather similar while the unloading steps are very different. The FQ exhibits very high elastic recovery while the BMG has a low one. (\textbf{Right}) ITSPM of one of the four tests is plotted for each material. Altitudes below 10 nm are masked in order to emphasize the shape and size of the residual piling-up. There is no visible residual piling-up on FQ, a low altitude circular one on WG and a higher altitude one with summits on the faces on BMG. These observations justify the choice of the three selected materials as they cover all possible contact geometries.}
\label{fig:figure_3}
\end{center}
\end{figure}

The tested materials (see Table \ref{tab:samples}) are chosen in order to cover a very wide range of contact geometries, from sinking-in (FQ), intermediate behavior (WG), and to piling-up (BMG). Glasses are chosen over metallic alloys because they exhibit negligible creep for temperatures well below the glass transition temperature, no visible size effect and are very homogeneous and isotropic in the test conditions. The FQ and WG samples are tested as received (please note the WG sample was test on it's "air" side) whereas the BMG sample is polished. Nanoindentation testing is performed using a commercial Hysitron TI950 triboindenter. During each test, the load is increased up to $P_{max} = 10$ mN  with a constant loading rate $ d P / d t = 5 \; \times \; 10^{-5}$ N/s. The load is then held for 10 s and relieved with a constant unloading rate  $ d P / d t = -1 \; \times \; 10^{-4}$ N/s. Four tests are performed on each sample and each residual imprint is scanned with the built-in ITSPM device with an applied normal force of 2 $\muup$N as summarized in Figure \ref{fig:figure_3}. Tests are load controlled and the maximum load is set to $P_{max} = 10$ mN. The true contact area $A_c$ is not known as in the case of the numerical simulations. It is then estimated through the Sneddon's Eq. \ref{eq:sneddon} \cite{Sneddon1965, ISI:000222316100002} and is called $A_{c,SN}$. Young's moduli $E$, and the Poisson's ratios $\nu$ of each sample are known prior to indentation testing from the literature or from ultrasonic echography measurements (\textit{cf.} Table \ref{tab:samples}). Recalling that $S$ is the initial unloading contact stiffness (cf. Figure \ref{fig:figure_1}), we have: 

\begin{equation}
A_{c,SN} = \frac{\pi}{4} \frac{S^2}{\beta^2 E_{eq}^2} \qquad \mbox{where: } \left\lbrace\begin{split}
E_{eq} = \frac{E}{1-\nu^2}\\
\beta = 1.05
\end{split}\right.
\label{eq:sneddon}
\end{equation}

\section{Methodology review}

\subsection{Direct methods}
Direct methods rely on the sole load vs. penetration curve $(P,h)$ to determine the contact height $h_c$  using equations given in Table \ref{tab:direct_methods}. Let us recall that $h_c < h$ in the case of sinking-in (as seen in Fig. \ref{fig:figure_2}) and $h_c>h$ for piling-up. Three direct methods are investigated in this paper : 
\begin{description}
\item[DN] The Doerner and Nix method \cite{JMR:7931088} was one of the first to be published (along with similar work done by Bulyshev \textit{et al.} \cite{Bulychev1975}) and it provided the basic relationships later improved by the two other methods.
\item [OP] The Oliver and Pharr method \cite{op1,ISI:000222316100002} is an all-purpose method that is widely used in the literature, commercial software and standards. The main drawback of this method is that it cannot take piling-up into account.
\item [LO] The Loubet method \cite{ISI:A1984SQ59800006} is an alternative to the OP method, especially for materials exhibiting piling-up.
\end{description}

Regardless of the chosen method, the value of $h_c$ is used to compute the value of the contact area $A_c$ thanks to the Indenter Area Function $A_c(h_c)$ (IAF). The IAF depends on the theoretical shape of the indenter as well and on its actual defects which are measured during a calibration procedure. Different tip calibration methods are used in the literature: 
\begin{itemize}
\item Measurement either of the indenter geometry or the imprint geometry made on soft materials for multiple loads using AFM or other microscopy techniques \cite{pethica1983}. 
\item The IAF introduced by Oliver and Pharr \cite{op1, ISI:000222316100002} requires a calibration procedure on a reference material using only the $(P,h)$ curve:

\begin{equation}
%A_{c,OP}(h_c) =\sum_{n=0}^8 C_n h_c^{2-n} 
A_{c,OP}(h_c) =C_0 h_c^2 + C_1 h_c + C_2 h_c^{1/2} +  C_3 h_c^{1/4} + \ldots + C_8 h_c^{1/128}  
\label{eq:hc_op}
\end{equation}

Where the $(C_i)_{0\leq i \leq 8}$ factors are fitting coefficients obtained from a calibration procedure on fused quartz. For a given indenter, the value of the $C_i$ coefficients depend on the penetration depth range used for the calibration procedure. In the case of a perfect modified Berkovich tip, $C_0 = 24.5$ and $C_{i>0} =0$. 

\item The method introduced by Loubet (see \cite{ISI:A1984SQ59800006}) :

\begin{equation}
A_{c,LO}(h) = k \left( h_c + \Delta h\right)^2
\label{eq:hc_loubet}
\end{equation}

It is assumed that the only origin of the defects is tip blunting. Then, $k$ comes from the indenter's theoretical shape ($k=24.5$ here) and $\Delta h$ is the offset caused by the tip defect and is calibrated using a linear fit made on the upper portion of the $(\sqrt P,h)$ curve. This procedure can be performed on any material exhibiting neither significant creep nor size effect, typically fused quartz. This method is intrinsically very efficient when the penetration is high compared to $\Delta h$. 

\end{itemize}

In the experimental benchmark, all tests are performed at $h_{max} \geq 250$ nm using a diamond modified Berkovich tip that exhibits a truncated length $\Delta h = 17.8\pm1.74$ nm\footnote{This value is the average value of $\Delta h$ on all the twelve tests performed on the three samples used in the experimental benchmark and the error is the represented $\pm$ one standard deviation.}. Theses values were calibrated on the FQ sample. As a consequence, the IAF introduced by Loubet is used on every direct method. By contrast, numerical simulations use a perfect tip so that the IAF is $A_{c}(h) = 24.5 h^2$.

\subsection{Proposed Method (PM)}
\label{sec:method}

\begin{figure}[p]
\begin{center}
\includegraphics[width = 1.\textwidth]{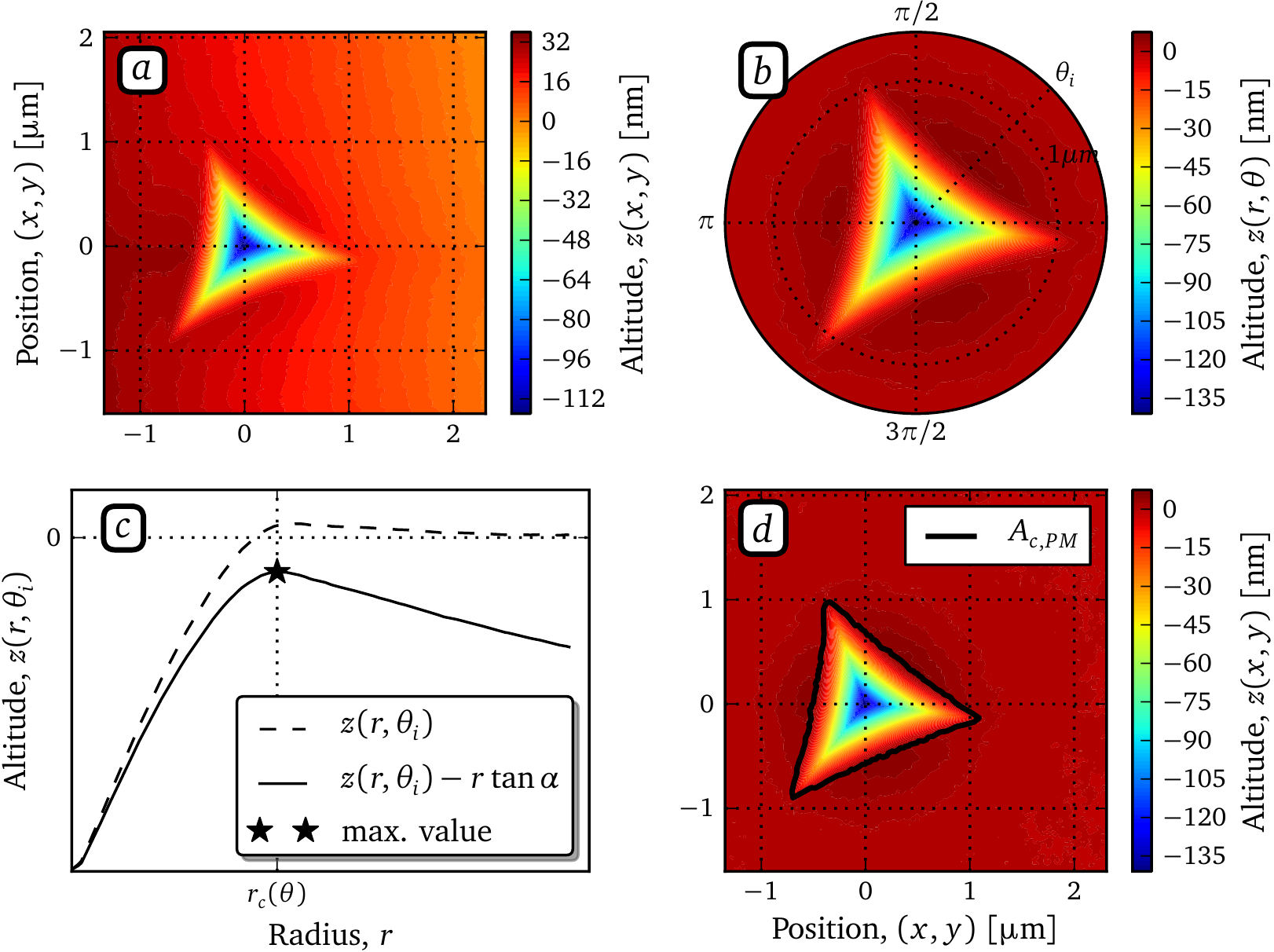}
\caption{Key steps of the proposed method exemplified on a Berkovich indent performed on window glass with $P_{max} = 10\;mN$. (\textbf{a}) Raw ITSPM image on which the tilt has to be corrected using a linear least squares fit on each individual scan with a circular mask around the indented region. (\textbf{b}) An half cross section at a given angle $\theta_i$ is extracted. (\textbf{c}) The half cross section $z(r, \theta_i)$ is rotated by a tilt angle $\alpha = 2.5$\degr~ in order to isolate more efficiently the edge of the contact zone. (\textbf{d}) The process is repeated for all required values of $\theta_i$ and the whole contact zone is extracted (in black).}
\label{fig:figure_4}
\end{center}
\end{figure}

SPM imaging grants access to a mapping of the altitude of the residual imprint. It is assumed that the surface of the sample is initially plane and remains unaffected far from the residual imprint. This plane is extracted from the raw image using a disk shaped mask centered on the imprint and a scan by scan linear fit on the remaining zone. It is considered as the reference surface and is subtracted to the raw image to remove the tilt of the initial surface. Under maximum load, the contact contour can exhibit either sinking-in and piling-up. In the first case, its altitude is decreased \textit{vis-à-vis} the reference surface. This behavior is typical of high yield strength materials such as fused quartz. On the opposite, piling-up occurs when the the increased and is typically triggered by unconfined plastic flow around the indenter as usually observed on low strain hardening metallic alloys. When the indenter is not axisymmetric (as it is the case for pyramids), both sinking-in and piling-up may occur simultaneously. For example, pyramidal indenters can produce piling-up on their faces and sinking on their edges (or no piling-up to the least). During the unloading step, the whole contact contour is pushed upward with only minor radial displacement. A residual piling-up may form even if sinking-in initially occurred under maximum load. We now focus on a half cross-section of the imprint starting at the bottom of the imprint and formulate two assumptions:
\begin{enumerate}
\item The highest point of any half cross-section is the summit of the residual piling-up.
\item The summit of the residual piling-up indicates the position of the contact contour.
\end{enumerate}
As a consequence, the highest point of the cross-section gives the radial position of the contact zone boundary. However, from an experimental point of view, the roughness of the sample's surface may make the true position of the residual piling-up's summit unclear. This issue is particularly true for materials exhibiting high levels of sinking-in such as fused quartz but also for most materials along the edges of pyramidal indenters. In order to limit the effect of surface roughness on the radial localization of the contact zone boundary, each profile is slightly rotated by a small angle value $\alpha$ along an axis perpendicular to the cross-section plane and running through the bottom of the imprint (\textit{i. e.} the lowest point of the profile). The whole contact contour is then determined by repeating the process in all directions ($\theta = 0$\degr~to $360$\degr) then the contact area $A_{c,PM}$ is calculated. A graphical representation of key steps of the method is made in Figure \ref{fig:figure_4}.

\noindent[PLEASE INSERT FIGURE 4 HERE]

\begin{figure}[p]
\begin{center}
\includegraphics[width = 1.\textwidth]{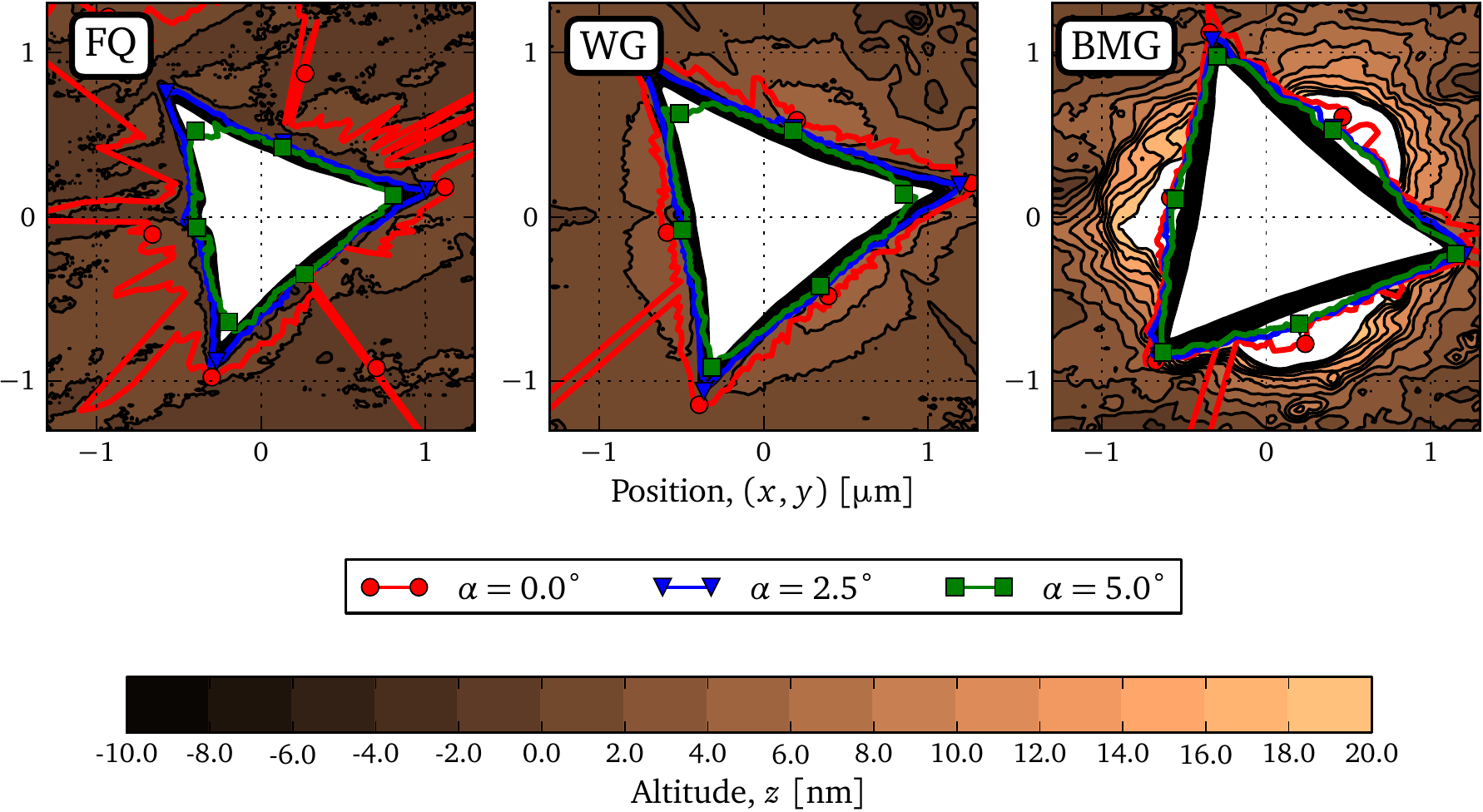}
\caption{\textcolor{blue}{Contact contours produced by the proposed method on the three residual imprints measured on each samples composing the benchmark. The imprints were produced using the experimental protocol described in the section \ref{subsec:expe_testing} and the Fig. \ref{fig:figure_3}. Three values of the tilt angle $\alpha$ are investigated on each imprint: no tilt ($\alpha = 0$\degr), the proposed value ($\alpha = 2.5$\degr) and last higher one ($\alpha = 5$\degr).}}
\label{fig:figure_8}
\end{center}
\end{figure}

\textcolor{blue}{
The optimum tilt value of the tilt angle $\alpha$ is chosen in order to deal with three potential artifacts detailed below and emphasized using three experimental residual imprints in Fig. \ref{fig:figure_8}:
\begin{enumerate}
\item Materials exhibiting low or no residual piling-up such as FQ cannot be treated with the proposed method when $\alpha = 0$\degr ~because the positions the highest points of each the extracted half cross sections are driven by the roughness. In this case, positive values of the tilt angle ($\alpha \geq 2.5$\degr) fully address the issue.
\item The surface roughness of the sample affects the accuracy of the method, especially on materials that show no residual piling-up because there is a competition between the is a competition between piling-up and roughness in the highest point identification (see FQ and BMG on Fig. \ref{fig:figure_8}. Again, positive values of the tilt angle ($\alpha \geq 2.5$\degr) solve the problem by lowering the roughness peaks located outside the imprint as visible.
\item A particular artifact is visible in the directions of the indenter's edges. Indeed, the residual piling-up is always very low in these directions because the displacement field tends to push the material towards the faces of the indenter. The attack angle between the edges and the sample's surface is also very low\footnote{ The angle is $13$\degr~ in the case a of modified Berkovich tip.} As a consequence, a high value of the tilt angle such as $\alpha \geq 5$\degr~ creates artifacts in this zone as clearly visible on FQ and WG. 
\end{enumerate}
Given the three above points, the optimum value of the tilt angle is found to be $\alpha = 2.5$\degr . The three values of the tilt angle have also been tested in on the numerical simulations. Even if the overall effect of the tilt angle is lower in this case because the simulations are two dimensional and also because the numerical samples have no roughness, only the first issue is observed and the optimum value is the same the one found experimentally.
}
\section{Results and discussion}
\label{sec:results}

\subsection{FEM benchmark results}
\begin{figure}[p]
\begin{center}
\includegraphics[width = 1.\textwidth]{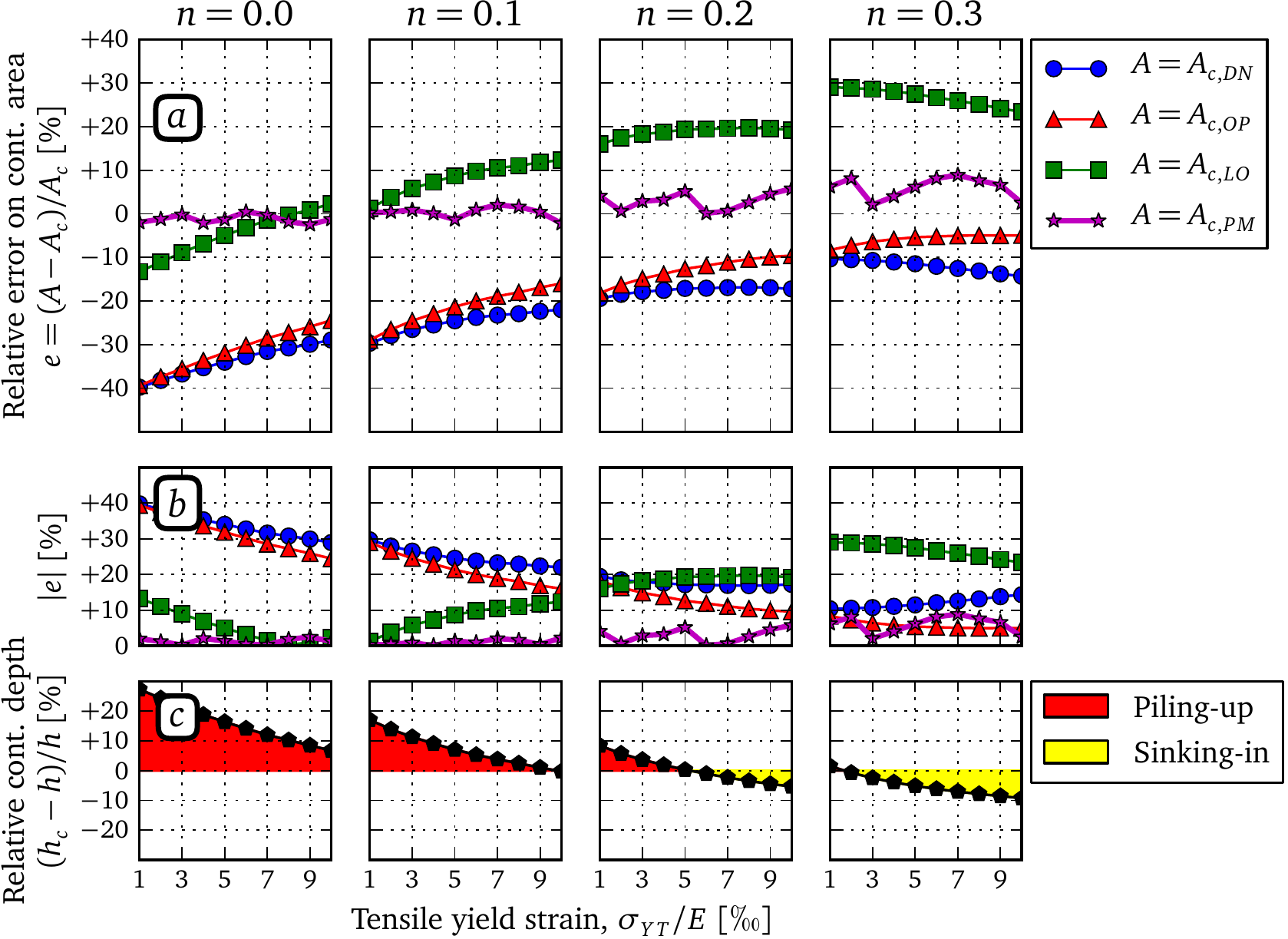}
\caption{Benchmark results of three direct methods and of the Proposed Method (PM) in the case of CE1. Different combinations of the tensile yield strain $\sigma_{YT}/E$ and the hardening exponent $n$ are investigated (see Table \ref{tab:hollomon_params}). On each simulation, the true projected contact area ($A_c$) computed by FEM, the contact areas estimated from the three direct methods ($A_{c,DN}$, $A_{c,OP}$ and $A_{c,LO}$) and the contact area from the proposed method ($A_{c,PM}$) are calculated. (\textbf{a}) The relative error $e$ between the true projected contact area and its four estimations is plotted. Each plot represents a different value of the hardening exponent $n$. (\textbf{b}) The absolute value of the the relative error $|e|$ is represented in order to emphasize the accuracy of each method (see Table \ref{tab:num_bench}). (\textbf{c}) The contact depth $h_c$ stands for the axial distance between the edge of the contact zone and the the summit of the indenter. The relative difference between the contact depth and the penetration $h$ is plotted to indicate where the piling-up occurs ($(h_c-h)/h > 0$) and when sinking-in occurs ($(h_c-h)/h < 0$). This subplot helps in understanding the relationships between the occurrence of piling-up and the accuracy of a given method.}
\label{fig:figure_5}
\end{center}
\end{figure}

\begin{figure}[p]
\begin{center}
\includegraphics[width = 1.\textwidth]{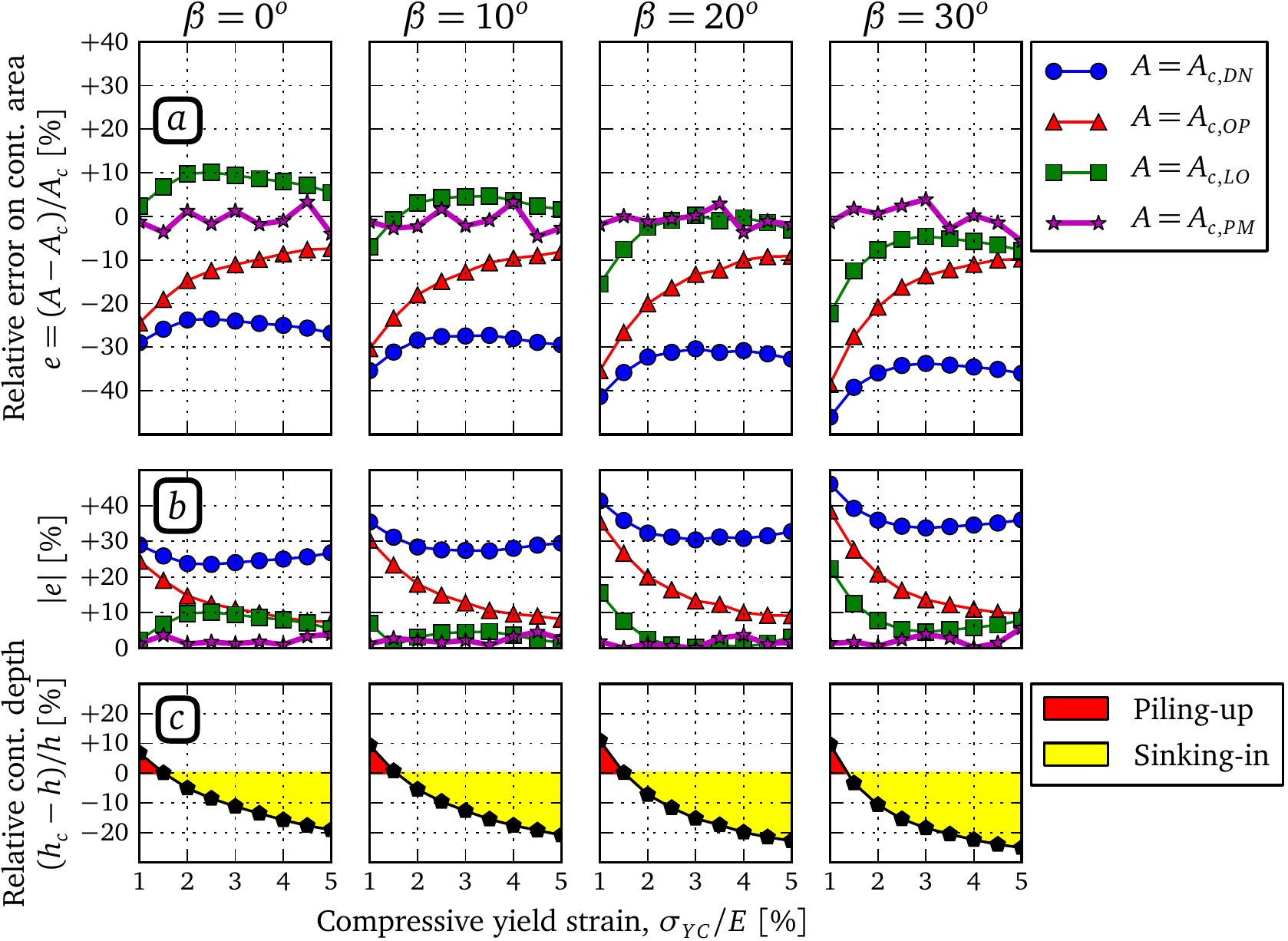}
\caption{Benchmark results of 3 direct methods and of the Proposed Method (PM) in the case of CE2. See Figure \ref{fig:figure_5} for the complete description of the structure of the figure.}
\label{fig:figure_6}
\end{center}
\end{figure}

The four methods are confronted on the numerical benchmark and their ability to accurately compute the contact area $A_c$ is challenged. The results focus on the relative error $e$ between $A_c$ and the contact area predicted by each method. Please note that a $10\%$ relative error on the contact area means roughly a $\approx 10\%$ relative error on hardness but only $\approx 5\%$ on the elastic modulus (see Eq. \ref{eq:sneddon}). The results are plotted in Figures \ref{fig:figure_5} and \ref{fig:figure_6} for constitutive equations CE1 and CE2 respectively and a summary of the key statistics is given in Table \ref{tab:num_bench}. For Fig. \ref{fig:figure_5} and \ref{fig:figure_6}, \textbf{(a)} and \textbf{(b)} represent the relative error and its absolute value respectively. The later was chose to emphasize the magnitude of the error while \textbf{(c)} indicates whether piling-up or sinking-in occurs. It is also chosen to measure the success rate of the methods through their ability to match $A_c$  within a $\pm 10\%$ error. This value was chosen since even if challenging, it is still realistic from an experimental point of view. These data can be discussed individually for each method: 
\begin{itemize}

\item The DN method systematically tends to underestimate $A_c$ regardless of the type of constitutive equation and of the occurrence of piling-up or sinking-in. The magnitude of the relative error is the highest among the four tested methods. This lack of accuracy can be put into perspective by recalling that the DN method states that the contact between the indenter and the sample behaves as if the indenter was a flat punch during the first stages of the unloading process. This approach was later proved to be too restrictive by Oliver and Pharr \cite{op1} who improved it by taking into account the actual shape of the indenter through the $\epsilon$ coefficient. As $\epsilon < 1$ in the case of the modified Berkovich tip, the value of the contact depth $h_c$ is systematically increased (see Table \ref{tab:direct_methods}). 

\item As stated above, the OP method drastically improves the overall performances of the DN method. However, its error level depends strongly on the type of contact behavior (\textit{i. e.} piling-up or sinking-in) and the mechanical properties of the tested material. Typically, it performs well for the CE1 law when the strain hardening exponent $n$ verifies $n>0.2$. It also performs well (relative errors below $10\%$) on materials exhibiting very high yield strains (higher than $4\%)$ in the case of CE2. The main drawback of the method is its intrinsic inability to cope with piling-up since $h_c/h$ can never be higher than 1. This is particularly visible for low values of the strain hardening exponent (CE1: $n\leq 0.1$) and with CE2 when the compressive yield strain $\sigma_C/E$ is lower than $3\%$. The OP method has a low success rate (see Table \ref{tab:num_bench})  but this tendency has to be mitigated by the fact that it is very efficient for a large number of metallic alloys, which can be described by CE1 and exhibit moderate values of hardening exponents. 
 
\item The LO method allows $h_c/h > 1$ values and is then recommended for piling-up materials; it is overall very efficient with CE2 type materials. The drawback is that it tends to overestimate the contact area when sinking-in occurs; this is particularly true in the case of CE1 with moderate to high hardening exponents ($n \geq 0.1$). These observations are in agreement with the results of Cheng  and Cheng \cite{cheng22} regarding the influence of piling-up and sinking-in on the direct estimation of the contact depth.  The success rate of this method is the highest among the direct methods and it is clearly the best available direct method for CE2 type materials and for low hardening CE1 type materials.

\item The proposed method exhibits a $100\%$ success rate (with the $\pm 10\; \%$ relative error target) and an average absolute relative error of $2.5 \%$. The error level remains stable regardless of both the type of constitutive equation and its parameters. This result highlights the fact that when experimentally possible, the use of such a \textit{post mortem} method will improve drastically the overall error level of the contact area measurement.
\end{itemize}

\subsection{Experimental benchmark results}

\begin{figure}[p]
\begin{center}
\includegraphics[width = 1.\textwidth]{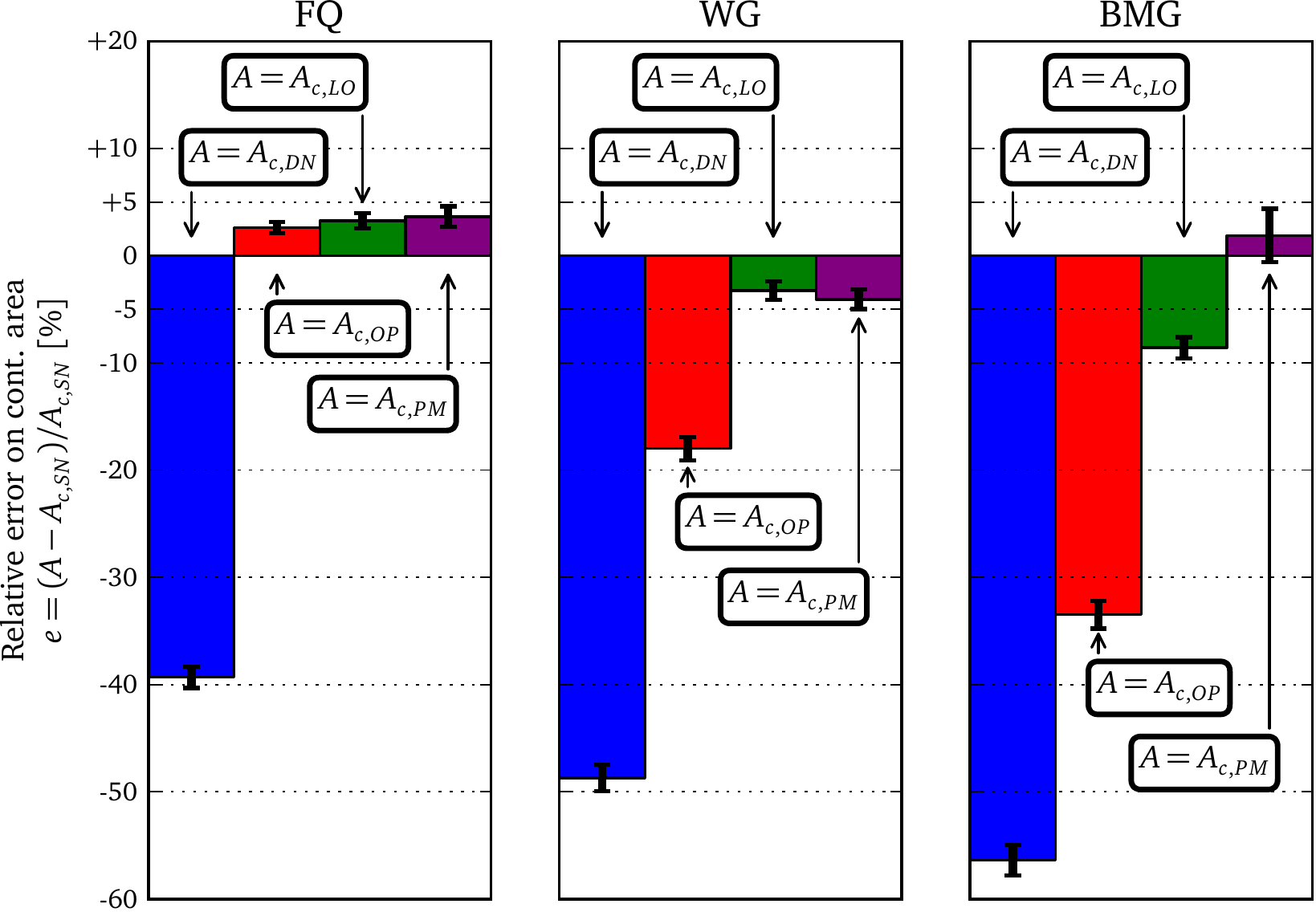}
\caption{Experimental confrontation of the three direct methods and the Proposed Method (PM) on the experimental benchmark. The relative error between the true projected contact area computed by Eq. \ref{eq:sneddon} and each method is plotted. Each subplot is dedicated to one sample and each bar within each plot refers to one of the four methods. The error bars represent the standard deviation obtained for four tests performed on each samples.}
\label{fig:figure_7}
\end{center}
\end{figure}

The results of the experimental benchmark are represented in Fig. \ref{fig:figure_7}. The tendencies observed in the numerical benchmark are confirmed. The DN method systematically underestimates the contact area. The OP and LO methods perform well only on a given spectrum of contact behaviors: both methods give accurate results on the FQ sample; this is consistent with the fact that both of them were optimized to use this material as a reference. While the LO method also exhibits a low error level on the WG sample, the OP method leads to an unexpected high error level. It is supposed that even if the WG sample has a very high yield strain, it has no strain hardening mechanism and it is then out of the scope of the OP method. The BMG which exhibits a large residual piling-up obviously leads the OP method to underestimate drastically the contact area. The LO method performs better although it also underestimates the contact area. This later method systematically exhibits relative errors of $\pm 10\%$ while the method proposed in this paper is even more reliable with errors lower than $5\%$. We observe that the direct methods overall performances are better than in the case of the numerical benchmark. The contact friction, which is neglected in the numerical benchmark, may improve the accuracy of the direct methods without affecting the proposed method.

\subsection{Pros and cons}

Both benchmarks highlight the precision gap between the new method and the existing direct methods. However, the proposed method differs by nature from the three direct methods it is compared to. This section emphasizes the pros and cons of this method:
\begin{description}
\item[Disadvantages:] 
\begin{itemize}

\item The proposed method relies on SPM imaging of the residual imprint while direct methods do not. However, indentation devices tend to be equipped with ITSPM capability that can be used automatically in conjunction with the indentation testing itself with only a small increase in test duration. 
\item \textcolor{blue}{The method relies on the assumption that the imprint is unchanged between the end of the test and the imaging procedure. This means that the proposed method should not be applied to materials exhibiting time-dependent mechanical behavior such as polymers (see \cite{Chatel2012}) and even pure metals. In order to demonstrate this limitation, nanoindentation tests have been run an electropolished pure single crystal aluminum sample (Al) and a mechanically polished pure tungsten sample (W). Those two materials were chosen because of their low elastic anisotropy \cite{Hirth1982}. A loading function similar to the one used in \cite{op1} was used to limit the effect of creep on the unloading stiffness $S$. The method appears to work flawlessly on both samples as none of the three artifacts mentioned in part \ref{sec:method} are observed. However, the value of the contact area is systematically overestimated by a factor $\approx 1.6$ and $\approx 2.7$ respectively on the Al and W samples. The evolution of the residual imprint after the indentation test is made possible by creep caused by residual stress.}
\end{itemize}
\item[Advantages:] 

\begin{itemize}
\item The proposed method can be run automatically, it requires no adjustable parameters and is user independent. 
\item Sample holder and machine stiffness affect the measurement of the penetration into surface $h$ as well as the measured contact stiffness $S$ and, as a consequence, they also affect all direct methods. The value of the machine stiffness can be measured once and for all while the sample holder's stiffness may change between two samples and requires systematic calibration. This concern is particularly true in the case of small samples such as fibers as well as very hard materials (such as carbides). The contact area measurement provided by the proposed method does not rely on $h$ and is then insensitive to the effect of those spurious stiffness issues. Yet, let us note that while the value of the contact area $A_{c,PM}$ is unaffected by stiffness issues, the value of the contact stiffness $S$ is of course affected. As a consequence, the value of the hardness probed with the proposed method is free from any stiffness concern (as $H = P_{max}/A_c$) while the value of the reduced modulus $E^*$ still requires stiffness calibration (as $E^*\propto S/\sqrt{A_c}$).
\item The method does not require any tip calibration procedure and is compatible with all tip shapes. 
\item The method is unaffected by erroneous surface detection also because it does not rely on $h$.
\end{itemize}    
\end{description}

\section{Conclusion}

We have proposed a new procedure to estimate the indentation contact area based on the residual imprint observation using altitude images produced by SPM. This area is the key component of instrumented indentation testing for extracting mechanical properties such as hardness or elastic modulus. For the estimation of this contact area, the method has been confronted with three widely used direct methods. We have showed, by means of an experimental and numerical benchmark covering a large range of contact geometries and materials, that the proposed method is far more accurate than its direct counterparts regardless of the type of material. We have also discussed the fact that such \textit{post mortem} procedures are indeed more time consuming than direct methods; yet they are the future alternative to direct methods with the development of indentation tip scanning probe microscopy techniques. We have also emphasized the fact that this new method has numerous advantages: it can be automated, it is user independent, it is unaffected by stiffness issues and does not require any indenter calibration.

\section{Acknowledgements} 

The authors would like to thank the Brittany Region for its support through the CPER PRIN2TAN project and the European University of Brittany (UEB) through the BRESMAT RTR project. Vincent Keryvin also acknowledges the support of the UEB (EPT COMPDYNVER).

\bibliographystyle{ieeetr}
%\bibliography{/home/lcharleux/Documents/Recherche/Bibliography/Mendeley/bibtex/library.bib}

\newpage

\begin{table}[h!]
	\centering
		\begin{tabular}{cccc} 
		\hline
		\textbf{Parameter} & \textbf{Description} & \textbf{Range} & \textbf{Number} \\
		$\sigma_{YT}/E$ & Tensile yield strain & $0.001 \rightarrow 0.01$&  10\\
		$n$ & Strain hardening exponent & $0.0 \rightarrow 0.3$ &  4\\
		\hline
		\end{tabular}
	\caption{Simulated range of the dimensionless ratios for the numerical simulations using the constitutive equation CE1. The "Number" column stands for the number of values chosen as simulation inputs in the given range.}
	\label{tab:hollomon_params}
\end{table}

\begin{table}[h!]
	\centering
		\begin{tabular}{cccc} 
		\hline
		\textbf{Parameter} & \textbf{Description} & \textbf{Range} & \textbf{Number} \\
		$\sigma_{YC}/E$ & Compressive yield strain & $0.01 \rightarrow 0.05$ & 9\\
		$\beta$ & Friction angle & $0.0$\degr $\rightarrow 30$\degr & 4\\
		\hline
		\end{tabular}
	\caption{Simulated range of the dimensionless ratios for using the constitutive equation CE2. The "Number" column stands for the number of values chosen as simulation inputs in the given range.}
	\label{tab:dp_params}
\end{table}

\begin{table}[h!]
	\centering
		\begin{tabular}{cccccc} 
		\hline
		\textbf{Label} & $\boldsymbol E$ \textbf{(GPa)} & $\boldsymbol \nu$ & \textbf{Provider/Description}& \textbf{Ref.} \\
		FQ & $73$ & $0.16$ & Hysitron Inc. & \\
		WG  & $71.5$ & $0.23$ & Planilux, Saint Gobain Inc. & \cite{ISI:000241652300022} \\
		BMG  & $89.3$ & $0.363$ &  Zr$_{55}$Cu$_{35}$Al$_{10}$, 385\degre~ annealed & \cite{Yokoyama2008}\\
		%BMG2  & $94.8$ & $0.359$ & $Zr_{50}Cu_{40}Al_{10}$, $400^oC$ annealed \\
		\hline
		\end{tabular}
	\caption{Description of the three tested samples: a fused quartz sample (FQ), a window glass sample (WG) and a zirconium based BMG. Materials properties used in this article are detailed: Young's modulus $E$, Poisson's ratio $\nu$. Reference to the literature are provided when available. In the case of fused quartz, the properties are given by the provider and are in agreement with the literature.}
	\label{tab:samples}
\end{table}

\begin{table}[h!]
	\centering
		\begin{tabular}{ccccc} 
		\hline
		\textbf{Method} & \textbf{Contact depth} & \textbf{Coeff.} & \textbf{Ref.} \\
		DN & $h_{c, DN} = h_{max} - P_{max}/S$ & \o & \cite{JMR:7931088} \\
	    OP  & $h_{c, OP} = h_{max} - \epsilon P_{max}/S$ & $\epsilon = 0.72$ & \cite{ISI:000222316100002} \\
        LO & $h_{c, LO} = \alpha \left(h_{max} - P_{max}/S\right)$ & $\alpha = 1.2$ & \cite{ISI:A1984SQ59800006}\\
		\hline
		\end{tabular}
	\caption{Equations used to compute the indentation contact depth $h_c$ for the three direct methods used in this article.}
	\label{tab:direct_methods}
\end{table}

\begin{table}[h!]
	\centering
		\begin{tabular}{ccccc} 
		\hline
		\textbf{Method} & $\boldsymbol{\min(e)\; [\%]}$ & $\boldsymbol{\max(e)\;[\%]}$ & $\boldsymbol{\overline{|e|}\;[\%]}$ & $\boldsymbol{|e| \leq 10 \% \;[\%]}$\\
		DN & -46.1 & -10.4 & 26.3 & 0.0 \\
		OP & -39.4 & -4.9 & 16.8 & 30.2 \\
		LO & -22.3 & 29.1 & 10.5 & 60.5 \\
		PM & -5.7 & 8.9 & 2.5 & 100.0\\
        \hline		
		\end{tabular}
	\caption{Numerical benchmark statistics of the three direct methods and the Proposed Method (PM) on both CE1 and CE2. Please note that 40 simulations were run for CE1 and 36 for CE2. As a consequence, the weight of CE1 is slightly higher than the weight of CE2 in the statistics. Mathematical notations are: $e = (A - A_c)/A_c$ is the relative error on the contact area $A$ computed by each method relatively to the the true contact area $A_c$, $|e|$ is its absolute value and $\overline{|e|}$ the arithmetic mean value of its absolute value. The last row displays the success rate of each method which is the proportion of the simulations on which the relative error is in the $\pm 10 \%$ range.}
	\label{tab:num_bench}
\end{table}

\newpage
%\section*{Figures}

\end{document}